\documentclass[entropy,article,submit,moreauthors,pdftex]{mdpi} 

\usepackage{amsmath}
\usepackage{latexsym}
\usepackage{amssymb}

\firstpage{1} 
\pubvolume{1}
\issuenum{1}
\articlenumber{0}
\pubyear{2020}
\copyrightyear{2020}
\history{Received: date; Accepted: date; Published: date}
\preto{\abstractkeywords}{\nolinenumbers}
\Title{Evolution toward linguistic coherence in naming game with migrating agents}

\Author{Dorota Lipowska $^{1,\dagger,*}$\orcidA{} and Adam Lipowski $^{2,\dagger}$}

\AuthorNames{Adam Lipowski, Dorota Lipowska}

\address{%
$^{1}$ \quad Faculty of Modern Languages and Literature, Adam Mickiewicz University in Pozna\'n, Poland; lipowska@amu.edu.pl\\
$^{2}$ \quad Faculty of Physics, Adam Mickiewicz University in Pozna\'n, Poland; lipowski@amu.edu.pl
}

\corres{Correspondence: lipowska@amu.edu.pl;}

\firstnote{These authors contributed equally to this work.}
\abstract{
As an integral part of our culture and way of life, language is intricately related to migrations of people. 
To understand  whether and how migration shapes language formation processes we examine the dynamics of the naming game with migrating agents. (i)
When all agents may migrate, the dynamics generates an effective surface tension, which drives the coarsening. Such a behaviour is  very robust and appears for a wide range of densities of agents and their migration rates. (ii) However, 
when only multilingual agents are allowed to migrate, monolingual islands are typically formed. In such a case, when the migration rate is sufficiently large, the majority of  agents acquire a common language, which spontaneously emerges with no indication of the  surface-tension driven coarsening. A relatively slow coarsening that takes place in a dense static population is very fragile, and most likely, an arbitrarily small migration rate can divert the system toward quick formation of monolingual islands. Our work shows that migration influences language formation processes but additional details like density, or mobility of agents are needed to specify more precisely this influence.
}

\begin{document}
%%%%%%%%%%%%%%%%%%%%%%%%%%%%%%%%%%%%%%%%%%
Although there is a multitude of factors that shape our language, including culture, politics, economy, geography or technology, the most important one is the mutual interactions  between multiple language users. It is thus tempting to examine language formation and its evolution using multi-agent models and statistical mechanics methodology \cite{loreto2011,loreto,blythe,patriarca}. A valuable insight into language emergence \cite{ellis}, death \cite{abrams}, its diversification \cite{serva}, importance \cite{ronen2014links}, or appearance of grammar or linguistic categories \cite{puglisi} proves that such an approach is indeed promising.

An important factor, which affects various aspects of our life, is migration. This process may mix as well as separate human communities and language formation processes should be thus strongly inﬂuenced by it \cite{routledge,williams}. Moreover, some modern trends, related mainly with globalization, most likely increase people’s migrations \cite{czaika}. 
It would be desirable to examine models that take into account both linguistic interactions and migration, and thus gain some understanding of how related are these two processes. A possible candidate is a suitably extended naming game, which proved to be useful in the studies of various aspects of the emergence of linguistic coherence \cite{steels,baron2016}. One of the questions, which may be addressed, is how coarsening, which in the naming game is known to be basically similar to the Ising model, is affected by migration.  Such a similarity is related to the fact that the dynamics of both models are driven by a surface tension \cite{bray}, even though in the naming game it is rather an effective surface tension \cite{dallastaepl2008}. Let us notice that recent studies indicate that the dynamics of English dialects evolution is also driven by  a certain effective surface tension \cite{burridge2018,burridge2020}. It would be desirable to check to what extent the surface-tension driven dynamics of the naming game is robust with respect to migration of agents. 

An interesting multi-agent model, where migration plays an important role, was proposed some time ago by Schelling \cite{schelling}. In his model, an agent is relocated if the number of its neighbours with the same orientation (an opinion or race) as the agent is below a certain threshold. Numerous versions of the Schelling model show that the phase separation (formation of a ghetto) appears to be a very robust feature of the dynamics \cite{schelling1}. Let us notice that in the Schelling model,  only those agents for which certain conditions are satisfied may migrate. Linguistic factors as, for example, acquiring a new language, might also influence our ability or willingness to relocate. 

In the present paper, we examine the naming game with migration and address some of the above issues. In particular, we examine whether an effective surface tension persists in such systems. We also examine the implications of a state-dependent migration but, of course, we are aware that its applicability to real social systems is limited.  Our simulations show that the state-dependent diffusion usually leads to spatial segregation but when sufficiently strong, it can trigger formation of the dominant language in the entire system.

\section{Model}

In our model, we consider a population of agents placed on a square lattice of linear size~$N$ (with periodic boundary conditions).  Each agent has its own inventory, which is a dynamically modified list of words.
Initially agents are uniformly distributed on the lattice with the density/probability $\rho$ (double occupancy excluded).  
The dynamics of our model combines the lattice gas random migration with the so-called minimal version of the naming game~\cite{steelsbaron,liplipmigrating2017}. 
 
More specifically, in an elementary step, an agent and one of its (four) neighbouring sites are randomly selected. With probability $d$, the agent migrates to the selected neighbouring site provided that the chosen site is empty and some additional conditions (dependent on the number of words in agent's inventory) are met. With probability $1-d$ and provided that the selected neighbouring site is occupied, the chosen agent becomes the Speaker, its neighbour becomes the Hearer, and they play the naming game: 
\begin{itemize}
\item Speaker selects a word randomly from its inventory (or invents a new word if its inventory is empty) and transmits it to Hearer. To invent a word all agents have at their disposal $M$ different words and one of them is selected randomly.
\item If Hearer has the transmitted word in its inventory, the interaction is a success and both players maintain only the transmitted word in their inventories.
\item If Hearer does not have the transmitted word in its inventory, the interaction is a failure and Hearer updates its inventory by adding this word to it.
\end{itemize}

The unit of time ($t=1$) is defined as $\rho N^2$ elementary steps, which corresponds to a single (on average) update of each agent.  Agents may have in their in ventories at most $M\geq 2$ different words but the coarsening dynamics of models with $M=2$ and $M>2$ is to some extent similar. Indeed, the evolution toward one of its $M$~absorbing states is driven by the effective surface tension \cite{castello2006,loreto2011} and analogous similarities are between the Ising and Potts models \cite{grest1988domain}. In the following, we will refer to words communicated by agents also as languages.

Recently, we have already analysed a naming game model with migration, in which, however, the relocation depended on the language used by an agent \cite{liplipmigrating2017}. In this model, all agents were allowed to migrate (albeit with a language-dependent rate) and the main objective of this study was to demonstrate  a certain symmetry breaking induced by the difference in mobility. In the present paper, the state-dependent mobility depends on the number of languages known by an agent and not on the particular language used by the agent. Some other systems with migrating agents but with different ordering dynamics (the voter model) were also analysed \cite{baronsatorras}. The emergence of consensus in the population was also examined in the case of agents moving in a continuous space \cite{baronchelli2012consensus} as well as in some robotic swarms \cite{cambier2017group}. It would be certainly interesting to replace the local migration of our agents with a possiblity of longer distance steps, as for example in some dynamical models defined on spatial networks \cite{gao2015bootstrap}.

Let us also emphasize that migration often refers to the mass movement of people in a certain direction or location, as e.g., in China during the Qing dynasty \cite{lee1991population}. To model such phenomena, a random walk of our agents would have to be considerably modified. 

%%%%%%%%%%%%%%%%%%%%%%%%%%%%%%%%
%%%%%%%%%%%%%%%%%%%%%%%%%%%%%%%%
\section{Results}
Naming game typically evolves toward a linguistic consensus state where all agents have only a single word in their repositories and thus every communication attempt results in a success.  
Before reaching such a state, monolingual domains are formed and their coarsening, driven by an effective surface tension, leads eventually to the emergence of a  linguistic consensus. The surface tension is known to drive the dynamics of many other models as, e.g., the Ising or Potts models \cite{bray}, and its absence as, e.g., in the voter model \cite{hinrichsen}, results in much different dynamics. In surface-tension driven dynamics, the correlation length can be related to the total length of domains' boundaries, which can be easily extracted from the model configuration. In the naming game, a competition between languages, which takes place at  domains'  boundaries, implies that such interfacial agents are typically bi-(or more)lingual.  Their concentration can be easily measured numerically and, being related to the total length of domains' boundaries, it determines the correlation length in the system.

\subsection{state-independent migration}
\begin{figure}
\includegraphics[width=\columnwidth]{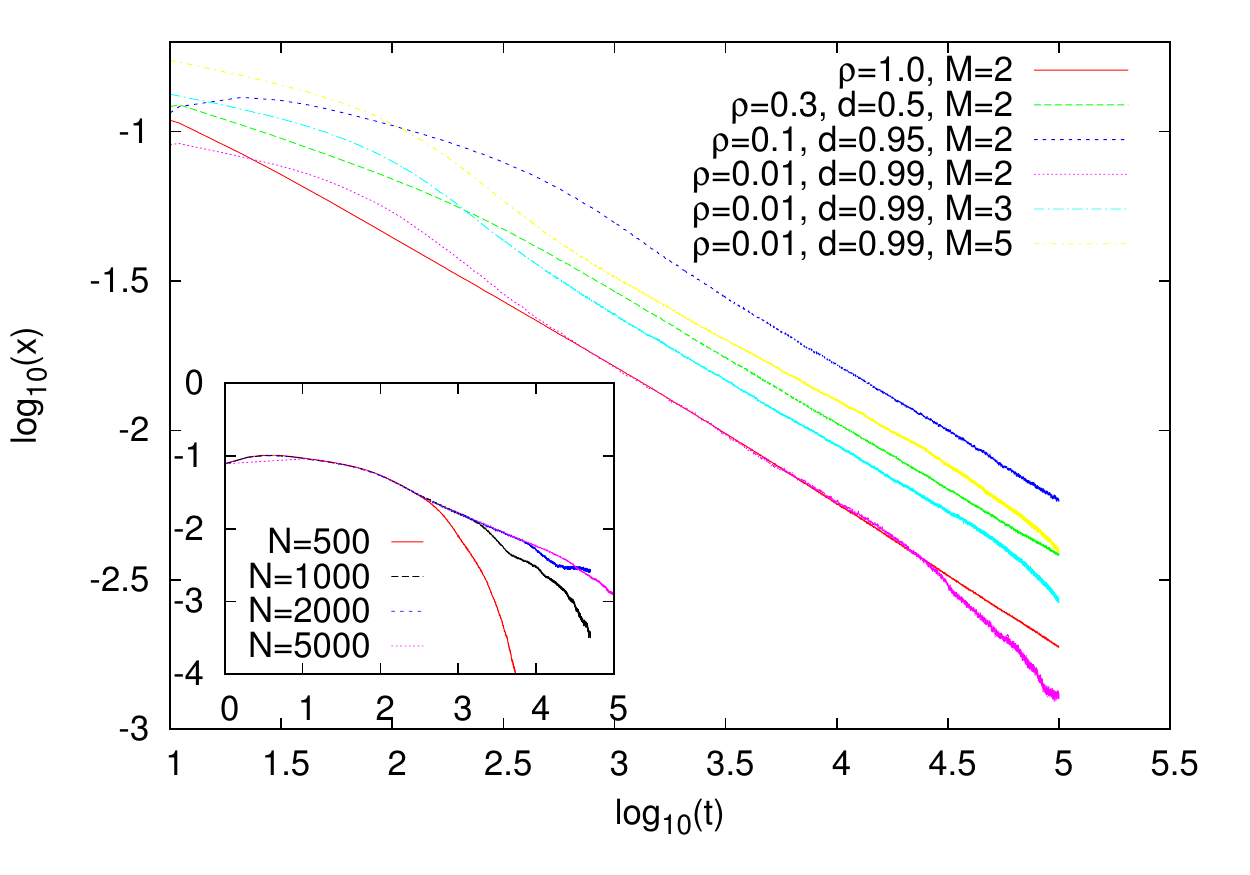}
\vspace{-8mm}
\caption{The time dependence of the fraction $x$ of bilingual agents for each set of parameters averaged over 100 independent runs. For $\rho\geq 0.1$, we present the results obtained for the system size $N=2000$ and we checked  that it was sufficiently large to avoid noticeable finite-size effects at the examined time scale. For $\rho=0.01$, we present the results for $N=5000$ and some late-time bending is still noticeable. The finite size effects  for $\rho=0.01$, $d=0.99$, and $M=2$  are illustrated in the inset. For $N\geq 1000$, a transient slowdown in the decay of~$x$ is seen, which we attribute to the formation of stripe-like structures \cite{lippref}. For increasing $N$, the influence of such stripes on the fraction~$x$ seems to diminish and a power-law decay sets in.}
\label{time}
\end{figure}

The case of our main concern is such that only agents with two or more words in their inventories  have the ability to migrate. However, first we report some results for the case, where all agents are able to migrate. We made simulations for several values of $\rho$ and $d$, and we measured the fraction $x$ of agents with two or more words in their inventories. The results of our simulations are presented in 
Fig.~\ref{time}. Most of our simulations were made for $M=2$  but some results for $M=3$ and $M=5$  show a similar behaviour. For $\rho=1$, when the lattice is fully occupied by agents and thus they have no space to migrate, $x$ shows a power law decay $x\sim t^{-\alpha}$. From our data we estimate $\alpha\approx 0.45(2)$, which agrees with some previous studies on the naming game \cite{baron2006,lippref} or related models of opinion formation \cite{dallasta2008,castello2006}. For $\rho<1$ and in presence of migration ($d>0$), the fraction~$x$ of multilingual agents seems to exhibit nearly the same  asymptotic decay. This is even the case when the lattice is sparsely covered with agents ($\rho=0.01$) that are engaged much more often in migration than in the naming game ($d=0.99$). Let us notice that in the Ising model with a nonconservative dynamics, the length of the interface is related to the excess energy above the ground state energy, and is known to decay as $t^{-1/2}$ (and that easily translates into $\sim t^{1/2}$ increase of the correlation length). The decay of $x$ observed in the naming game is very similar and most likely it is related to a certain effective surface tension generated in this kind of models  \cite{dallastaepl2008}.

Our simulations show that the surface-tension driven dynamics in the naming game is very robust with respect to the concentration of agents and migration rate.  There are some reasons to believe that some other factors will not change qualitative features of the dynamics of such systems either. Indeed, we have recently shown that the surface-tension driven dynamics is restored in the voter model (which is known to have much different dynamics and no surface tension \cite{hinrichsen}) with only a small fraction of sites evolving according to the Ising heat-bath dynamics \cite{liplipfer2017}. It is thus plausible that even in heterogeneous systems, where some of our naming-game agents would be replaced with agents with dynamics similar to the voter model, the system would still exhibit surface-tension characteristics. Provided that the naming game mimics to some extent real linguistic interactions, such a strong robustness could suggest  that a surface-tension driven dynamics could operate  in (real) processes responsible for the evolution of natural languages, which certainly complies with some recent analysis indicating that an effective surface tension seems to shape patterns of dialect changes \cite{burridge2018,burridge2020}.
%%%%%%%%%%%%%%%%%%%%%%%%%%%%%%%%%%
%%%%%%%%%%%%%%%%%%%%%%%%%%%%%%%
\subsection{state-dependent migration}
Agents with two or more words in their inventories, whose fraction is denoted by~$x$, are typically located at the interface of monolingual domains and thus may be considered as multilingual. Since the ability (or willingness, or need) to migrate is not necessarily homogenous in the population, we would like to examine the case when only such multilingual agents may migrate.

Such a state-dependent migration resembles the Schelling model of ghetto formation, where an agent is relocated if the number of its neighbours of the same (as the agent) orientation is too small \cite{schelling,schelling1}. Let us notice that in the Schelling model the orientation of an agent is fixed during the evolution of the model, which is not the case in our model. In such an analogy, our model could be considered as driven by a nonconservative dynamics.
Simulations of such a model reveal that both the density $\rho$ and migration rate $d$ influence the dynamics of the model and its final state. We made simulations in the low- ($\rho=0.3$) and high-density ($\rho=0.8$) regimes.  In these 
regimes there are some qualitative differences in the behaviour of the model that we describe in the following.

\subsubsection{$\rho=0.3$}

\begin{figure}
\includegraphics[width=\columnwidth]{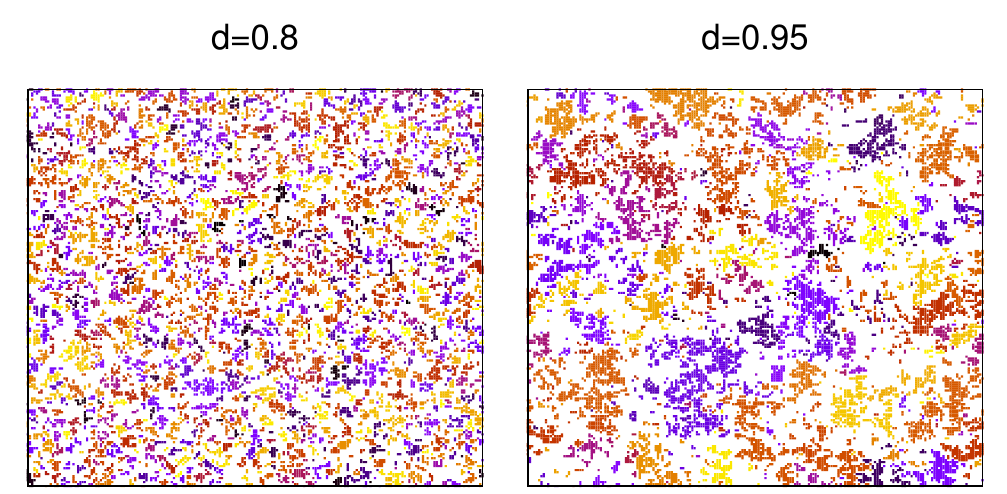}
\vspace{0mm}
\caption{The spatial distribution of agents and languages they use after $t = 10^3$ for $d=0.8$ and after $t = 10^4$  for $d=0.95$.  After such transients, all agents form separated clusters and become monolingual and thus immobile (see Fig.~\ref{time03}).  Simulations were made for $N=200$ and $\rho=0.3$; different colours correspond to different languages.}
\label{confp03d08095}
\end{figure}

In this subsection we describe our results in a low-density regime, and as a representative value we have chosen $\rho=0.3$ and $M=1000$.  Simulations show that in this case, after a relatively short transient, agents rearrange and form monolingual clusters separated from each other (Fig.~\ref{confp03d08095}).

%%%%%%%%%%%%%%%%%%%%%%%%%%%%%%%%%%
%%%%%%%%%%%%%%%%%%%%%%%%%%%%%%%

\begin{figure}
\includegraphics[width=\columnwidth]{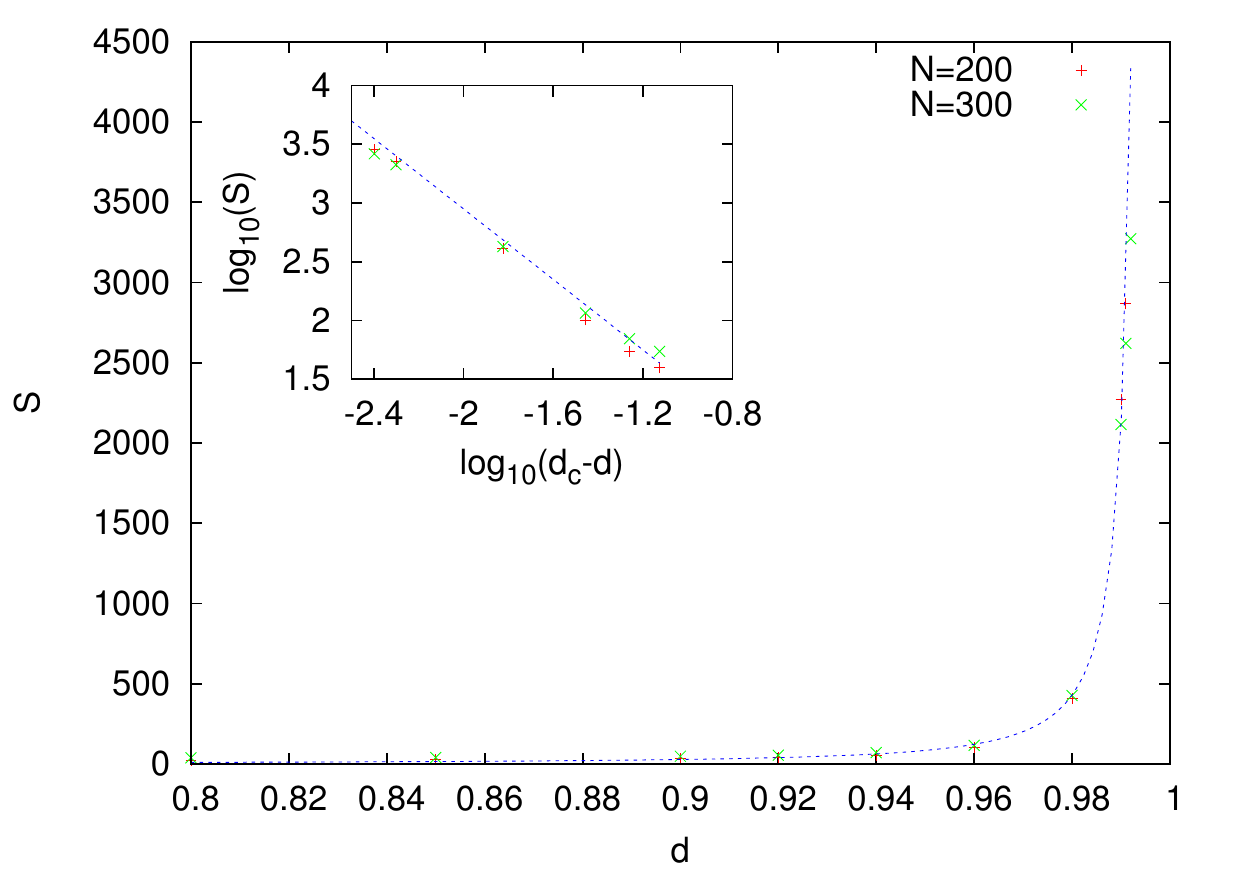}
\vspace{-5mm}
\caption{The average size of the surviving languages as a function of the mobility $d$. The inset presents the same data on a logarithmic scale. The least-square fit  shows that numerical data follow a power-law divergence $S\sim (d_c-d)^{-\gamma}$ with $d=d_c=0.995(1)$ and $\gamma=1.5(1)$ (dashed line).}
\label{avsize}
\end{figure}

The size of these clusters seems to increase with the migration rate $d$. Calculation of the average size of such clusters $S$ suggests that $S$ may diverge at the value $d=d_c$ close to but smaller than $d=1$ (Fig.~\ref{avsize}).

%%%%%%%%%%%%%%%%%%%%%%%%%%%%%%%%%%
%%%%%%%%%%%%%%%%%%%%%%%%%%%%%%%

\begin{figure}
\begin{center}
\includegraphics[width=0.8\columnwidth]{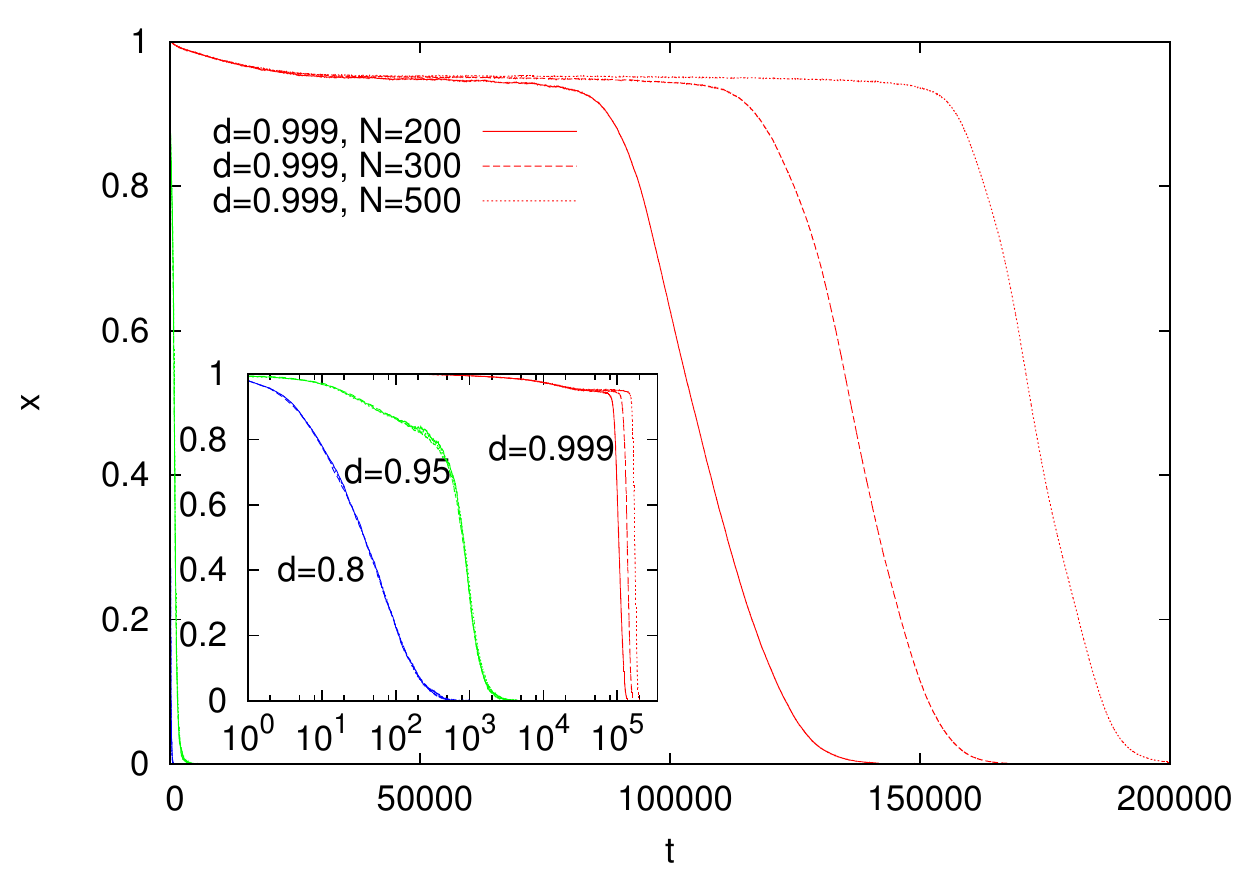}
\end{center}
\vspace{0mm}  
\caption{The time dependence of the fraction $x$ of mobile agents calculated for $d=0.999$ (red), 0.95 (green), and 0.8 (blue). Simulations were made for $\rho=0.3$ and system size $N=200$ (continuous lines), 300 (dashed), and 500 (dotted). The presented results are averages over 30 independent runs. For $d=0.8$ and 0.95, the decay of $x$ takes place on a short time scale and is nearly size independent.  The inset shows the same data but with a logarithmic time axis.}
\label{time03}
\end{figure}

 It is  interesting to examine the evolution of our model for $d_c<d<1$. In particular, for $d=0.999$ simulations show  that the fraction of mobile agents $x$ drops to 0 but at a time scale that shows a pronounced size dependence (Fig.~\ref{time03}). For $d=0.8$ and 0.95 no such dependence is observed (inset in Fig.~\ref{time03}).

%%%%%%%%%%%%%%%%%%%%%%%%%%%%%%%%%%
%%%%%%%%%%%%%%%%%%%%%%%%%%%%%%%

\begin{figure}
\includegraphics[width=\columnwidth]{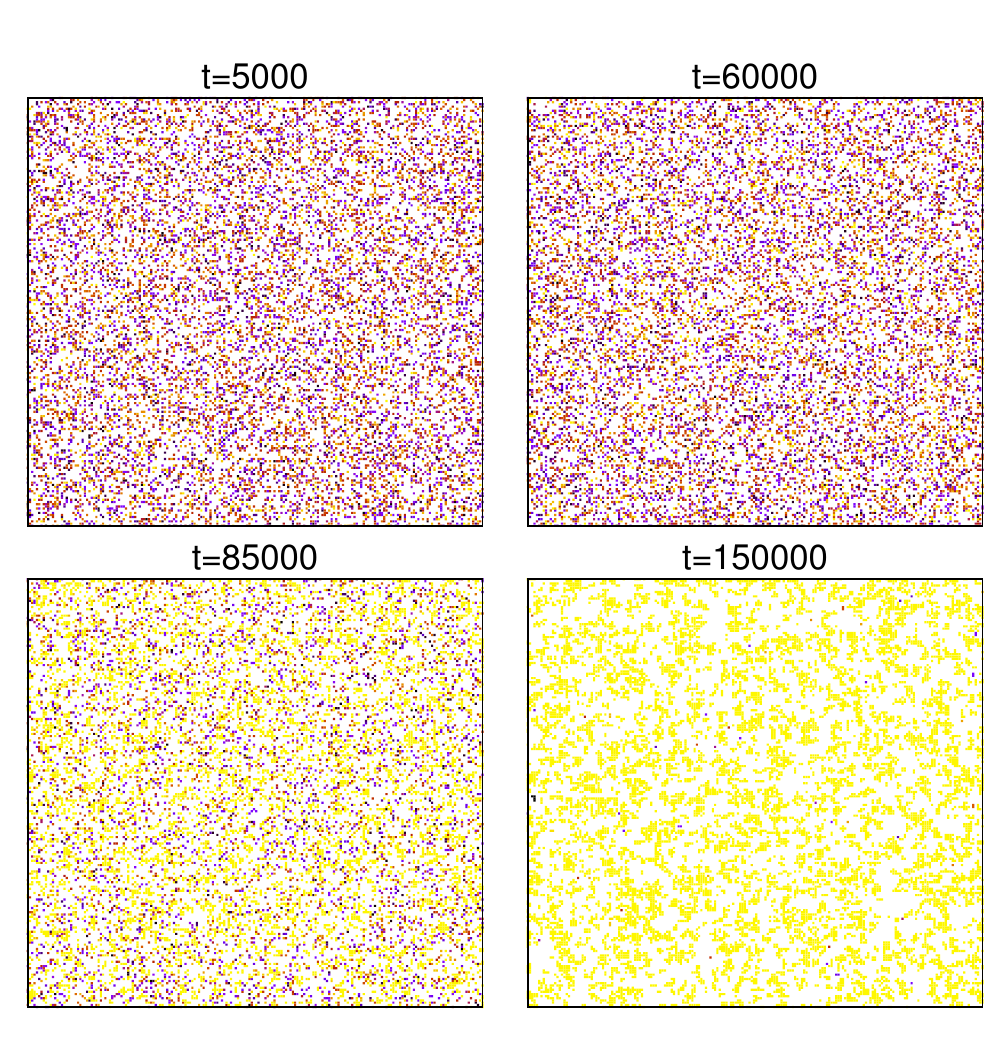}
\vspace{-5mm}
\caption{Snapshots of spatial distribution of agents and languages they use for $d=0.999$, $\rho=0.3$, and $N=200$. The dominant language emerges (around $t=85000$) without any indication of an Ising-like coarsening.}
\label{confp03d0999}
\end{figure}

According to Fig.~\ref{avsize}, for $d=0.999$ the average size of a cluster is infinite, which indicates formation of a dominant language that is used by  the majority of agents.
Visual analysis of the time evolution of the model for $d=0.999$ shows that in this case the dominant language emerges without any indication of the coarsening, which is so characteristic to the naming game and other surface-tension driven models (Fig.~\ref{confp03d0999}). It may suggest that a sufficiently strong state-dependent diffusion diminishes (to zero?) the surface tension and the dominant cluster emerges in the system similarly as in the (multi-state) voter model. The time scale of such transition, possibly diverging with the system size (Fig.~\ref{time03}), supports such an interpretation. Let us emphasize, however, that the estimated critical value $d_c$ is very close to~1. Although the size of the surviving languages (Fig.~\ref{avsize}) for both $N=200$ and $300$ seems to diverge at nearly the same value $d=d_c=0.995(1)$, we cannot exclude that this is actually the finite size effect and, in the thermodynamic  limit, $d_c=1$.
However, the true thermodynamic limit is perhaps not that important in the linguisitc contexts and the dynamics that we observe might be relevant and interesting also as a finite size effect. 
%%%%%%%%%%%%%%%%%%%%%%%%%%%%%%%%%%
%%%%%%%%%%%%%%%%%%%%%%%%%%%%%%%%%%

\subsubsection{$\rho=0.8$}
\begin{figure}
\begin{center}
\includegraphics[width=0.5\columnwidth]{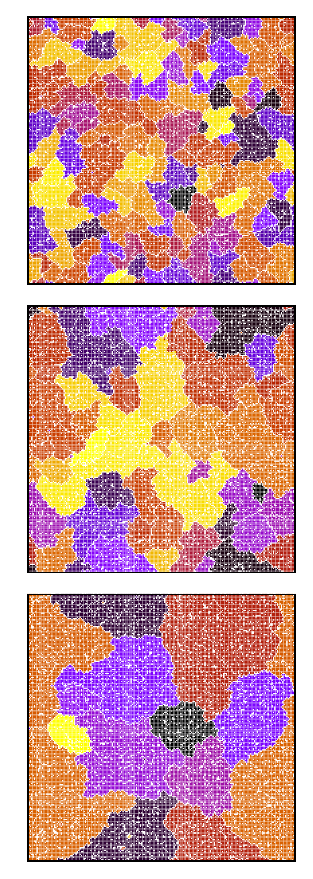}
\end{center}
\vspace{0mm}
\caption{The final spatial distributions of agents and languages they use for $\rho=0.8$, $N=200$, $M=1000$ and $d=0.1$ (top), $d=0.01$ (middle), $d=0.001$ (bottom).}
\label{confp08d001}
\end{figure}

In this subsection, we examine the behaviour of our model for larger density, namely for  $\rho=0.8$. Similarly to the $\rho=0.3$ case, migration coupled with the ordering mechanism of the naming game leads finally to the formation of monolingual islands separated by some empty spaces and examples of such structures are seen in  Fig.~\ref{confp08d001}. Let us notice that the decrease of the migration rate $d$ results in the increase of the size of islands, contrary to the $\rho=0.3$ case, where the size of islands was increasing with an increase of~$d$. However, configurations presented in Fig.~\ref{confp08d001} are obtained for small values of migration rate $d$. For larger $d$ (close to $d=1$),  a trend similar to that for $\rho=0.3$ will perhaps be observed. Since the system with $\rho=0.8$ is relatively dense, numerical investigation of such a regime might be difficult.

%%%%%%%%%%%%%%%%%%%%%%%%%%%%%%%%%%
\begin{figure}
\begin{center}
\includegraphics[width=0.8\columnwidth]{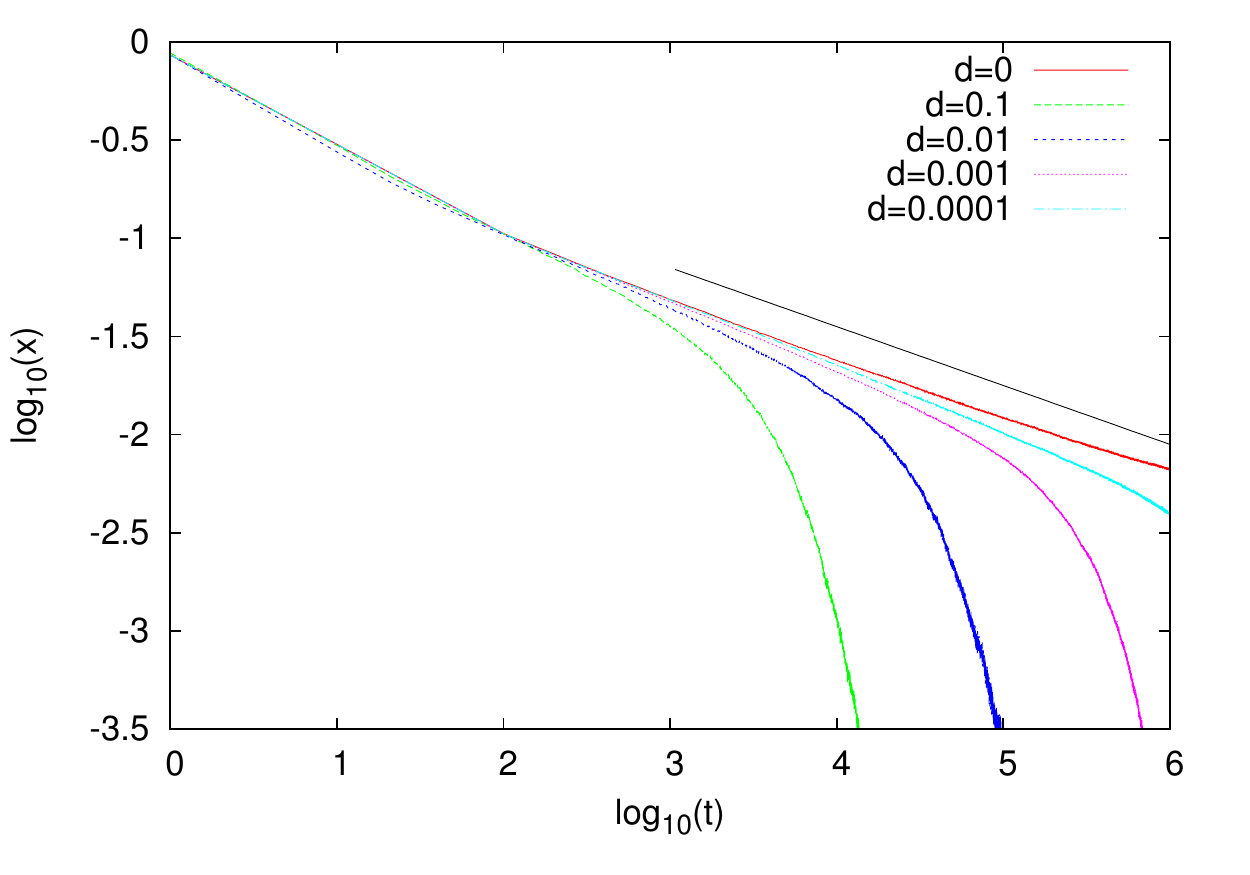}
\end{center}
\vspace{-8mm}
\caption{The time dependence of the fraction $x$ of bilingual agents calculated for $\rho=0.8$, $N=1000$ and several values of mobility $d$. The solid straight line has a slope corresponding to $x\sim t^{-0.3}$ and some bending of our data indicates that the asymptotic decay for $d=0$ might be even slower than that. The presented results are  averaged over 100 independent runs.}
\label{timep08}
\end{figure}

%%%%%%%%%%%%%%%%%%%%%%%%%%%%%%% 
For $N=1000$ and $M=1000$, we also calculated the time dependence of~$x$ (Fig.~\ref{timep08}). 
In the absence of migration ($d=0$), we observe a power-law-like decay of~$x$ but some bending of our data suggests that the estimated decay $\sim t^{-0.3}$ may  not be truly asymptotic. A slower decay, perhaps logarithmically slow, would be actually consistent with the coarsening of the Ising model on diluted lattices \cite{corberi}. Leaving aside the correct asymptotic form of the decay of~$x$, we would like to point out that the $d=0$ coarsening that  takes place in our model for $\rho=0.8$ is an expected feature assuming some analogy to the Ising model, with which the naming game seems to share some similarities. Indeed, above the site percolation threshold ($\rho_c \approx 0.5928$ \cite{stauffer,stauffer2}), the Ising model is expected to order ferromagnetically  at sufficiently low temperature with the accompanying power-law coarsening \cite{corberi2017coarsening, neda1994curie}, and we expect a similar behaviour in the naming game.
 Below such a threshold ($\rho<\rho_c$), our agents are located in finite clusters, which rather quickly become monolingual due to the naming game.

For $d>0$, we observe that $x$ shows a rapid (probably faster than the power-law) decay (Fig.~\ref{timep08}). We associate such a decay with the formation of monolingual islands (Fig.~\ref{confp08d001}).
What is in our opinion surprising is that even a very small migration rate~$d$ is sufficient to bring the system to such a multi-island configuration. Indeed, even for $d=10^{-4}$ the data seem to veer off the $d=0$ line. With $d$ decreasing, this deviation takes place at an increasing time scale, which is related to the formation of islands of an increasing size.

%%%%%%%%%%%%%%%%%%%%%%%%%%%%%%%%%%
%%%%%%%%%%%%%%%%%%%%%%%%%%%%%%%

\begin{figure}
\begin{center}
\includegraphics[width=0.6\columnwidth]{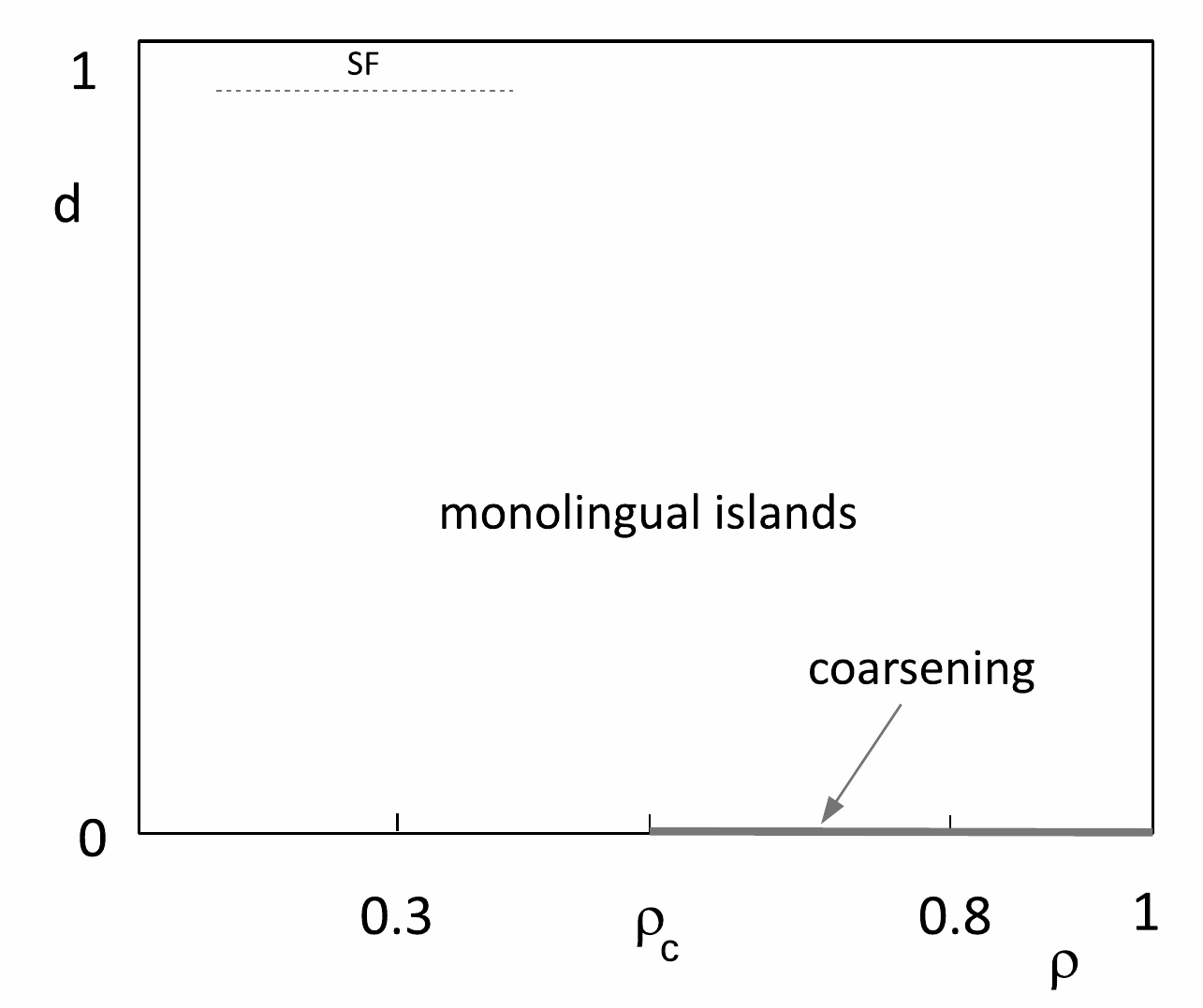}
\vspace{-0mm}
\caption{The $(\rho, d)$ phase diagram as inferred from our simulations. In the largest portion of the phase diagram the final state is quickly reached and is made of finite-size monolingual islands. Only for $d=0$ and $\rho>\rho_c$ we expect a power-law coarsening and for $\rho<1$ the coarsening, due to random dilution, is probably slower than $t^{1/2}$.  For $d$ very close to 1 we expect the regime where the dominant language is formed via a spontaneous fluctuation  ( SF, the dashed line is not in scale). }
\label{phase-diagram}
\end{center}
\end{figure}

The overall behaviour of our model for the state-dependent migration is presented in the phase diagram in 
Fig.\ref{phase-diagram}.

%%%%%%%%%%%%%%%%%%%%%%%%%%%%%%%%%%
%%%%%%%%%%%%%%%%%%%%%%%%%%%%%%%

\section{Conclusions}
In summary,  we examined how an ordering dynamics of the naming game is affected by migration. In the version where all agents are allowed to migrate, the coarsening of our model indicates the presence of an effective surface-tension. Such a behaviour is very robust with respect to the concentration of agents or their migration rate. Recently, we have shown that an effective surface tension appears in a heterogenous voter model with a small fraction of agents operating with the Ising heat-bath dynamics \cite{liplipfer2017}. Since the naming game shares some similarity with the Ising model, one may thus hope that an effective surface tension should be a generic feature of language formation models, which would be resilient against dilution, migration or dynamical heterogeneities.  This is very much in accord with some recent analysis, where the surface tension was shown to shape the English dialects evolution \cite{burridge2018,burridge2020}. It was also suggested \cite{burridge2020} that the diffusion of language users  may reduce the surface tension and shift the dynamics toward voter-like. Our simulations do not support such a behaviour, at least within the scope of our model. 

When only multilingual agents were allowed to migrate, we observed formation of monolingual islands. Such a state-dependent migration dynamics resembles that of the Schelling model and islands may be considered as analogues of ghettos that are typically formed in this model. Similarly to the Schelling model, the formation of islands is a robust feature of the dynamics and it takes place for small as well as large concentration of agents. Let us notice that, unlike the Schelling model, our agents might change their language and with this respect they are driven by nonconservative dynamics. Our simulations suggest that when state-dependent migration rate is sufficiently large, a certain language becomes dominant and spreads over the majority of agents. However, the transition toward such a linguistic coherence is not a surface-tension driven coarsening but rather a spontaneous fluctuation, similar perhaps to the transition in the voter model. The predicted migration-induced reduction of the surface tension \cite{burridge2020} would thus take place but only with  migration of multilingual agents. When the concentration of agents is above a site percolation threshold and the state-dependent migration is absent, agents form an infinite cluster and the naming game dynamics induces the coarsening albeit slower than on an undiluted lattice. Such a coarsening appears to be very fragile with respect to the state-dependent migration and most likely, an  arbitrarily small migration directs the dynamics toward formation of monolingual islands.

Migration is an important factor that should be taken into account in studying language formation models as well as some other agreement dynamics systems. 
It would be certainly desirable to develop alternative approaches that would allow for at least qualitative understanding of our results, which are based only on numerical simulations. Field-theory techniques based on the Fokker-Planck equation were used to analyse a related class of models, the so-called voter model with intermediate states \cite{dallasta2008},  and it would be interesting to develop a similar approach in the context of our models. However, taking into account migration of our particles is likely to result in a more complex field-theory description.   It would be also interesting to examine whether an effective surface tension appears also in reinforcement learning systems with migration~\cite{liplipplos} or in heterogeneous systems, where agents evolve with different kinds of dynamics. Elucidation of the role of the state-dependent migration in formation of a dominant language or in a high fragility of slow coarsening on a diluted lattice would be also desirable.

%%%%%%%%%%%%%%%%%%%%%%%%%%%%%%%%%%%%%%%%%%
\authorcontributions{conceptualization, D.L. ; methodology, A.L.; software, A.L. and D.L.; validation, A.L. and D.L.; investigation, A.L.  and D.L..; writing--review and editing, A.L. and D.L.; visualization, A.L. All authors have read and agreed to the published version of the manuscript.'', please turn to the  \href{http://img.mdpi.org/data/contributor-role-instruction.pdf}{CRediT taxonomy} for the term explanation. Authorship must be limited to those who have contributed substantially to the work reported.}

%%%%%%%%%%%%%%%%%%%%%%%%%%%%%%%%%%%%%%%%%%
\conflictsofinterest{The authors declare no conflict of interest.} 
%%%%%%%%%%%%%%%%%%%%%%%%%%%%%%%%%%%%%%%%%%
\reftitle{References}

%\externalbibliography{yes}

%\nocite{*}
%\bibliographystyle{mdpi}
%\bibliography{abbrev_titles,biblio}

%=====================================
% References, variant B: internal bibliography
%=====================================
%\begin{thebibliography}{999}
%% Reference 1
%\bibitem[Author1(year)]{ref-journal}
%Author1, T. The title of the cited article. {\em Journal Abbreviation} {\bf 2008}, {\em 10}, 142--149.
%% Reference 2
%\bibitem[Author2(year)]{ref-book}
%Author2, L. The title of the cited contribution. In {\em The Book Title}; Editor1, F., Editor2, A., Eds.; Publishing House: City, Country, 2007; pp. 32--58.
%\end{thebibliography}

%% for journal Sci
%\reviewreports{\\
%Reviewer 1 comments and authors’ response\\
%Reviewer 2 comments and authors’ response\\
%Reviewer 3 comments and authors’ response
%}

%%%%%%%%%%%%%%%%%%%%%%%%%%%%%%%%%%%%%%%%%%
\end{document}